\documentclass[aps,prd,twocolumn,superscriptaddress]{revtex4-2}
\usepackage{graphicx}  
\usepackage[export]{adjustbox}
\usepackage[caption=false]{subfig}
\usepackage{svg}
\usepackage{hyperref}  
\hypersetup{colorlinks=true, citecolor=blue, urlcolor=blue, linkcolor=blue}
\usepackage{dcolumn}   
\usepackage{bm}        
\usepackage{amssymb}   
\usepackage{amsmath}   
\usepackage{euscript}
\usepackage{verbatim}
\usepackage{setspace}
\usepackage{xcolor}
\usepackage{amsfonts}
\usepackage{braket}
\DeclareUnicodeCharacter{2212}{-}

\newcommand{\STO}{SrTiO$_3$}
\newcommand{\LAO}{LaAlO$_3$}
\newcommand{\alumina}{Al$_2$O$_3$}
\newcommand{\STOLAO}{\LAO \slash\STO{}}
\newcommand{\STOLAOalumina}{\alumina \slash\LAO\slash\STO{}}
\newcommand{\Tc}{T$_c$}

\allowdisplaybreaks

\begin{document}
\title{Enhanced Superconductivity in \STO-based Interfaces via Amorphous \alumina{} Capping}

\author{I. Silber}
\affiliation{School of Physics and Astronomy, Tel−Aviv University, Tel Aviv, 69978, Israel}
\author{A. Azulay}
\affiliation{Department of Materials Science and Engineering, The Iby and Aladar Fleischman Faculty of Engineering, Tel−Aviv University, Tel Aviv, 6997801, Israel}
\author{A. Basha}
\affiliation{Department of Materials Science and Engineering, The Iby and Aladar Fleischman Faculty of Engineering, Tel−Aviv University, Tel Aviv, 6997801, Israel}
\author{D. Ketchker}
\affiliation{School of Physics and Astronomy, Tel−Aviv University, Tel Aviv, 69978, Israel}
\author{M. Baskin}
\affiliation{The Andrew and Erna Viterbi Faculty of Electrical and Computer Engineering, Technion – Israel Institute of Technology, Haifa 32000-03, Israel }
\author{A. Yagoda}
\affiliation{School of Physics and Astronomy, Tel−Aviv University, Tel Aviv, 69978, Israel}
\author{L. Kornblum}
\affiliation{The Andrew and Erna Viterbi Faculty of Electrical and Computer Engineering, Technion – Israel Institute of Technology, Haifa 32000-03, Israel }
\author{A. Kohn}
\affiliation{Department of Materials Science and Engineering, The Iby and Aladar Fleischman Faculty of Engineering, Tel−Aviv University, Tel Aviv, 6997801, Israel}
\author{Y. Dagan}
\affiliation{School of Physics and Astronomy, Tel−Aviv University, Tel Aviv, 69978, Israel}

\begin{abstract}

\textbf{Oxide interfaces feature unique two-dimensional (2D) electronic systems with diverse electronic properties such as tunable spin-orbit interaction and superconductivity. Conductivity emerges in these interfaces when the thickness of an epitaxial polar layer surpasses a critical value, leading to charge transfer to the interface. Here, we show that depositing amorphous alumina on top of the polar oxide can reduce the critical thickness and enhance the superconducting properties for the (111) and the (100) \STO-based interfaces. A detailed transmission electron microscopy analysis reveals that the enhancement of the superconducting properties is linked to the expansion of the \LAO{} lattice in a direction perpendicular to the interface. We propose that the increase in the superconducting critical temperature, \Tc{}, is a result of epitaxial strain.}
\end{abstract}

\maketitle

\section{Introduction}
The phenomenon of superconductivity in semiconductors with an extremely low electron density remains a puzzle. The mystery lies in how these electrons, characterized by a substantial Coulomb repulsion and a limited density of states, can condense into a superconducting ground state. A hallmark example is doped \STO \cite{Schooley1964}, which superconducts at very low carrier densities (n $\approx$ 6$\times10^{17}$cm$^{-3}$, less than one electron per million unit cells) \cite{Lin2013}, and its superconducting mechanism is under debate \cite{Wolfle2018,Ruhman2019, Collignon2019,Gastiasoro2020}. Moreover, the carrier-temperature phase diagram shows superconductivity in a dome-shaped region \cite{Koonce1967,Binnig1980,Thiemann2018,Fauque2023}. This unusual behavior has also been observed in \STO-based interfaces \cite{Caviglia2008,Maniv2015}. However, the superconducting transition temperature \Tc{} in interfaces has always been smaller than the bulk counterpart \cite{Lin2014,Gariglio2016}.

Interfacial conductivity can occur when a layer of a polar oxide, such as \LAO, is deposited on an undoped \STO. When the polar layer surpasses a specific thickness, conductivity emerges \cite{Thiel2006}. The electron system is limited to a few unit cells near the interface \cite{Basletic2008, Yamada2014, Fister2014}. The straightforward explanation for the observed criticality in polar-layer-thickness is the necessity for a large enough electric field to initiate charge transfer to the interface \cite{Ohtomo2004,Pentcheva2009,Yu2014}. This critical thickness can be reduced by depositing a metallic \cite{Lesne2014,Vaz2017, Bisht2022} or another oxide \cite{Singh2018,Kwak2021} layer on top of the polar oxide. However, the decrease of the critical thickness usually does not improve the superconducting properties compared with the bare \STOLAO{} interface.

Here we show that by deposition of an amorphous alumina capping layer on top of an ultrathin epitaxial \LAO{}, the (111) \STOLAOalumina{} interface becomes conducting and superconducting with \Tc{} = 420mK, which is about 20\% higher than the uncapped ("bare") interface. Transmission electron microscopy study shows that the \LAO{} lattice expands perpendicular to the interface for this configuration, while it contracts for the sample with a thicker \LAO{} and no alumina capping. We also show a reduction of the critical \LAO{} thickness for the (100) interfaces but with \Tc{} = 365mK. We establish a correlation between \Tc{} in both interfaces and the expansion of the \LAO{} lattice, proposing that the increased superconducting critical temperature is a result of the strain imposed by the \LAO{} layer on the uppermost \STO{} layer, where conductivity and superconductivity reside.

\section{Sample design and structural analysis}
We fabricated a series of \STOLAOalumina{} samples with a constant amorphous alumina thickness of 10nm and varying \LAO{} thickness both in the [100] and [111] directions of the \STO{} substrate (see Table S1). We deliberately employed deposition conditions to prevent surface reduction of \STO. The \LAO{} thickness is measured in unit cells (UC) comprising LaO and AlO$_2$ layers when deposited along the [100] direction and in monolayers (ML) consisting of one layer of LaO$_3$ and one layer of Al when deposited along the [111] direction.

\begin{figure*}[ht] 
\includegraphics[width=0.75\linewidth]{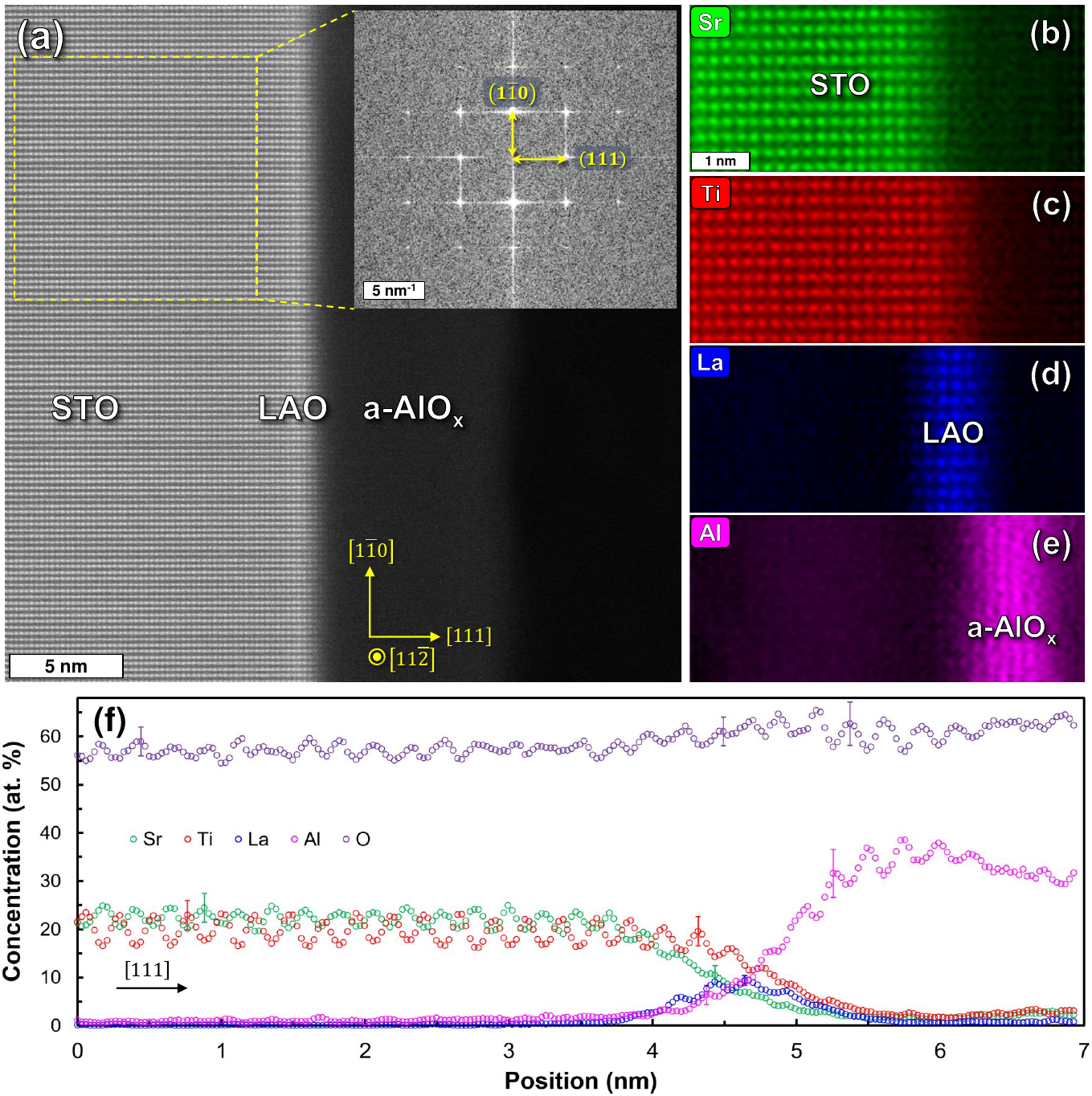}
\caption{(a) Cross-sectional high-angle annular dark-field scanning transmission electron microscope image showing atomic-column resolution of the sample comprising 3 \LAO{} monolayers covered by amorphous alumina, grown on a \STO{} (111) substrate and oriented to the [11$\bar{2}$] zone axis. The inset shows the power spectrum calculated from the region marked by the yellow square. (b)-(e) Elemental maps obtained by energy-dispersive X-ray spectroscopy (EDS) using net intensities of Sr L, Ti K, La L, and Al K. (f) Concentrations of Sr, Ti, La, and Al along the [111] direction, obtained by integration over the width of the images shown in (b)-(e) followed by Cliff-Lorimer k-factor quantification. Error bars represent typical compositional uncertainty in that region.}
\label{FIG1}
\end{figure*}

Figure \ref{FIG1}(a) shows a cross-sectional high-angle annular dark-field scanning transmission electron microscope (HAADF STEM) image of the 3 ML sample, revealing STO[111][1$\bar{1}$0]//LAO[111][1$\bar{1}$0] epitaxial relation, as expected (see Figure S1). The highlighted peaks in the power spectrum are attributed to reciprocal lattice vectors $g_{1\bar{1}0}$ and $g_{111}$, namely interplanar distances of $d_{110}$ $\sim$ 0.276 and $d_{111}$ $\sim$ 0.225nm in STO of the \textit{P}\textit{m}$\bar{3}$\textit{m} space group (\textit{a} = 0.3905nm \cite{Lytle1964}).   

Elemental maps obtained by energy-dispersive X-ray spectroscopy (EDS), representing net intensities of Sr L, Ti K, La L, and Al K, are shown in Figure \ref{FIG1}(b)-(e). These maps reveal Sr, Ti, La, and Al atomic columns viewed along the [11$\bar{2}$] zone axis, e.g., Sr and Ti alternating at (111) planes, as expected from our design of the \STOLAOalumina{} structure. Integration of the signal over the width of the images shown in Figure \ref{FIG1}(b)-(e) followed by elemental quantification is presented in Figure \ref{FIG1}(f). The oscillating peaks of Sr and Ti concentration suggest that the Sr to Ti atomic ratio is $\sim$ 1 in the \STO{} region. However, the gradual decrease in Sr content through the \LAO{} region indicates partial substitution of Sr for La, as observed previously \cite{Nakagawa2006,Willmott2007}. Furthermore, for the amorphous alumina region, around a 2:3 Al:O atomic ratio is observed. Similar results are obtained for the 14 ML sample, see Figure S2. We note that the oscillating peaks of Al concentration in the amorphous alumina region may stem from partial crystallization during EDS acquisition \cite{Liu1993}. Also, we note that the minor La and Al concentrations observed in the STO region are statistically insignificant (see Figure S3).
\section{results}
\subsection{Reduced Critical Thickness}
In bare \STOLAO{} interfaces, the critical thickness for conductivity is four unit cells of \LAO{} for the (100) interface \cite{Thiel2006} and nine monolayers for the (111) interface \cite{Herranz2012}. 
However, adding an alumina layer over \STOLAO{} interfaces decreases the critical thickness for conductivity.

\begin{figure}[ht] 
  \centering
\includegraphics[width=\linewidth]{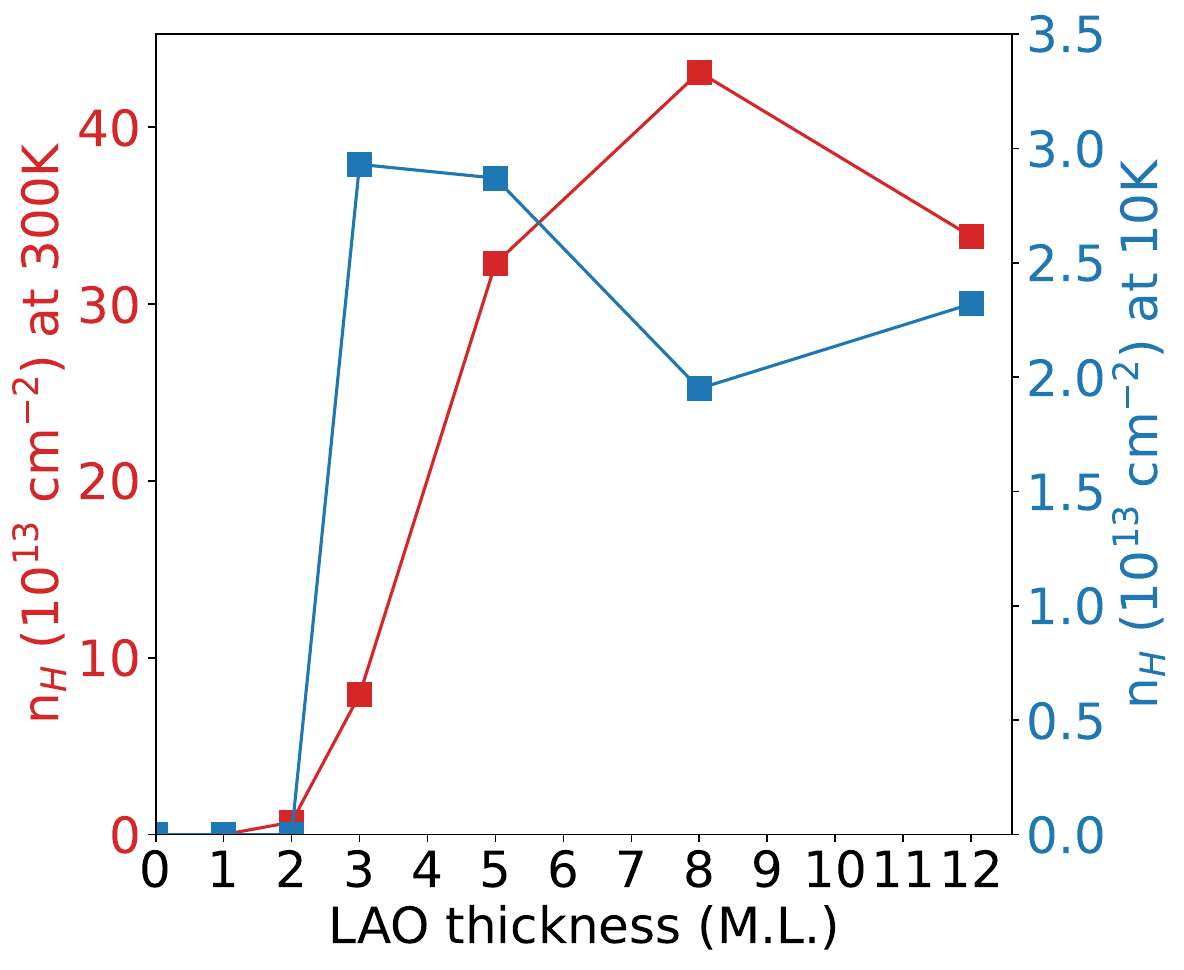}
\caption{Reduced critical thickness of  (111) \STOLAOalumina{} interface. The measured Hall number at 300K and 10K, as a function of the \LAO{} thickness. The  \LAO{} thickness controls whether the sample is insulating, metallic or superconducting.}
\label{HallTwoTemps}
\end{figure}

In the (100) interface, the threshold reduces to two UC of \LAO{}, while in the (111) \STOLAOalumina{} interface, it decreases to 3 ML of \LAO{}. This effect can be clearly seen in the Hall number shown in Fig. \ref{HallTwoTemps}. The variation of the Hall number with temperature is similar to previous studies \cite{Ryu2022}. 

We stress again that the \alumina/\STO{} interface is not conducting under our deposition parameters. The few \LAO{} layers are probably needed to induce a strong enough dipole moment on the surface of the \STO{} as suggested in Ref. \cite{Yu2014}. Our results showing reduction of the critical thickness in (111) \STOLAOalumina{} are in accordance with density functional theory calculations that predicted a reduction of the \LAO{} critical thickness when a neutral layer is added \cite{Cho2022}. 

\subsection{Enhanced Superconductivity}
\begin{figure}[ht] 
  \centering
\includegraphics[width=\linewidth]{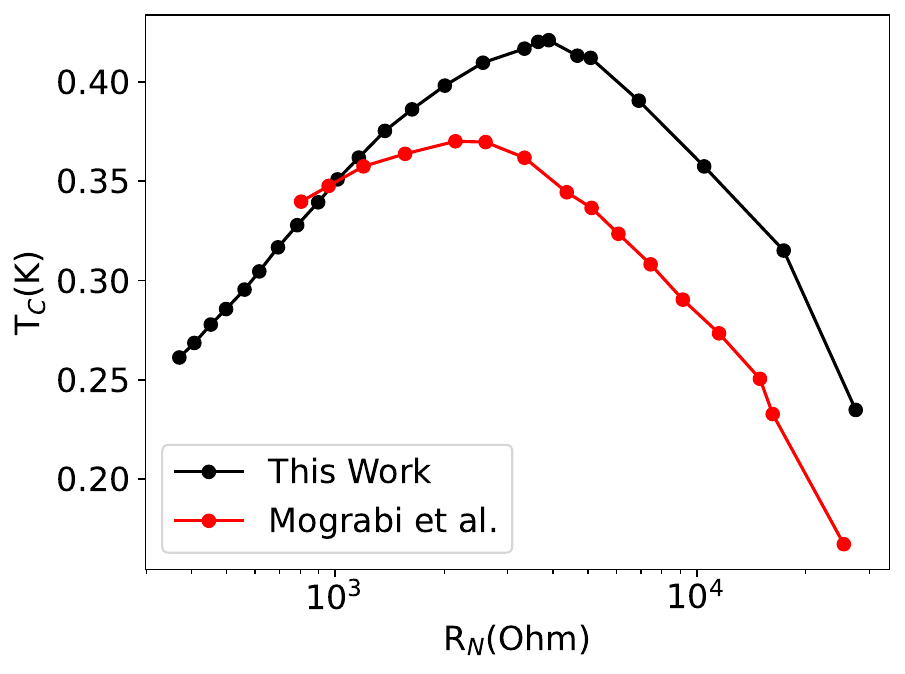}
\caption{Enhanced superconductivity with the alumina capping for the 3 ML \STOLAOalumina{}. The dependence of thr superconducting critical temperature versus sheet resistance (controlled by gate voltage) exhibits a dome-shaped line of the phase transition. A significant enhancement of the superconducting critical temperature is observed, comparing with a sample without the alumina capping (data for the bare (111) \STOLAO{} sample are taken from our previous study \cite{Mograbi2019}).}
\label{IA3Dome}
\end{figure}

Surprisingly, at the critical \LAO{} thickness of 3 ML, the (111) \STOLAOalumina{} has an enhanced superconducting \Tc{}, as seen in Figure \ref{IA3Dome}. This effect starkly contrasts with previous works, in which the reduction of the \LAO{} critical thickness resulted in similar or poorer superconducting properties relative to the bare \STOLAO{} interface. 

\begin{figure}[ht] 
  \centering
\includegraphics[width=\linewidth]{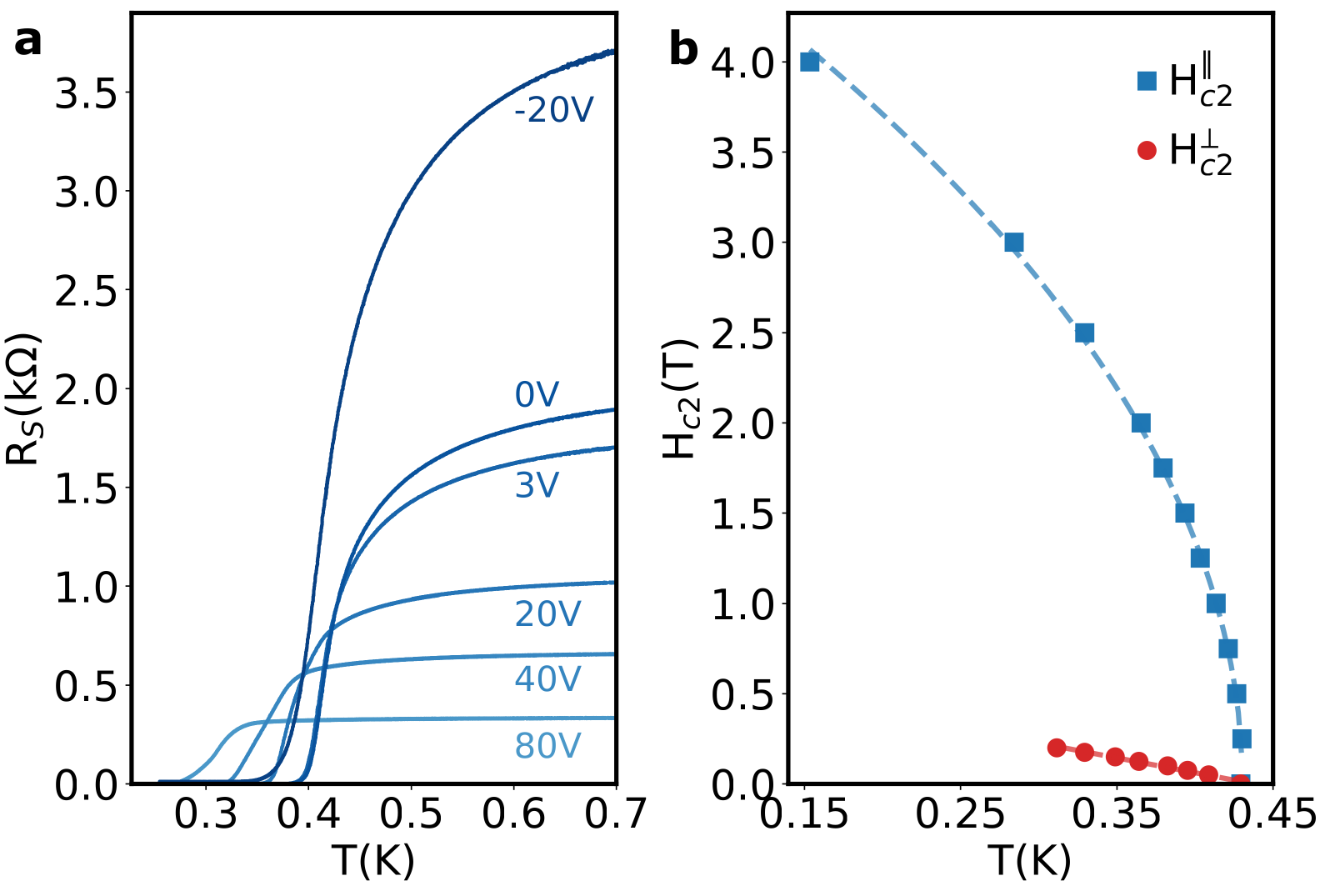}
\caption{Superconducting properties of the \STOLAOalumina{} interface for \LAO{} thickness of 3 monolayers. (a) The superconducting transition at various back-gate voltages, displaying the non-monotonic dependence of the transition temperature on the sample normal state resistance. (b) Critical field (zero gate voltage), when the field is applied in-plane or out-of-plane. The characteristic behavior of two-dimensional superconductivity is clearly demonstrated. Dashed lines are fits to square root for the parallel field configuration and linear for the perpendicular field.}
\label{FIG2}
\end{figure}

The superconducting state of the 3 ML \STOLAOalumina{} sample is two-dimensional in nature as inferred from the different temperature dependence of the critical field for parallel and perpendicular magnetic field orientations (Fig. \ref{FIG2}b). Furthermore, the gate tunability of the conducting and superconducting properties is a characteristic of a two-dimensional conductor (Fig. \ref{FIG2}a).

From our measurements, we find that \Tc{} = 420mK, the Ginzburg–Landau coherence length $\xi_{GL}$ = 21.4nm, and the upper bound for the superconducting layer thickness is 10nm. The length scales are comparable to those found in the uncapped (111) \STOLAO{} interface \cite{Rout2017}. In the [100] direction we observed a smaller 5\% enhancement of the critical temperature compared to the bare \STOLAO{} interface. See Table S1 and supplemental material for the superconducting properties of the (100) \STOLAOalumina{} samples. 

\subsection{Comparative Strain Mapping}
\begin{figure*}[ht] 
\includegraphics[width=0.75\linewidth]{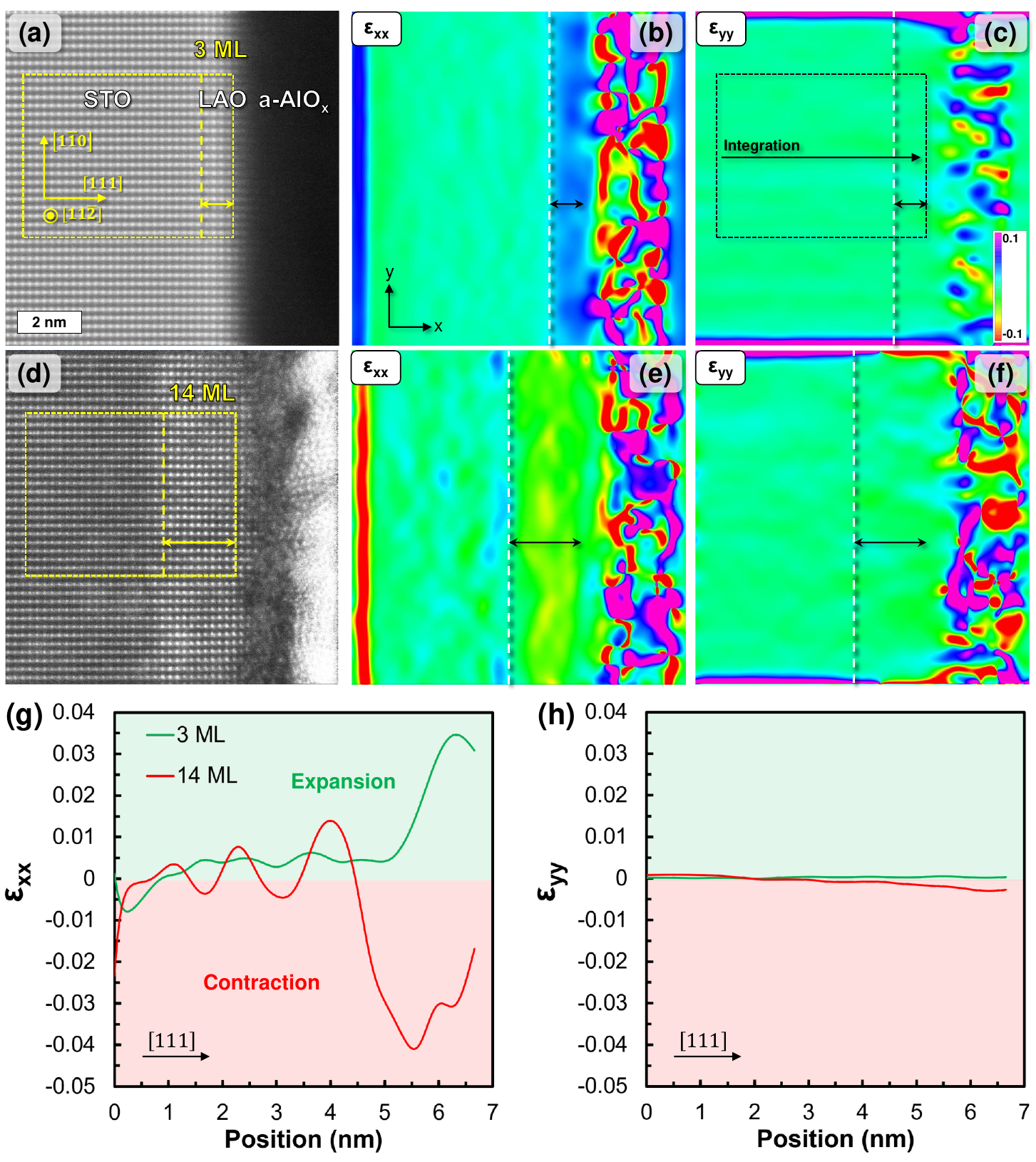}
\caption{Strain within the \STOLAOalumina{} and \STOLAO{} interfaces extracted from high-angle annular dark-field scanning transmission electron microscope (HAADF STEM) images across the [11$\bar{2}$] zone axis by the geometric phase analysis method. (a) and (d) HAADF STEM images of cross-sectional samples comprising 3 \LAO{} ML covered by amorphous alumina (3 ML) and 14 \LAO{} ML (14 ML), grown on a \STO{} substrate, respectively. (b)-(c) and (e)-(f) Strain maps corresponding to the [111] and [1$\bar{1}$0] directions ($\varepsilon_{xx}$ and $\varepsilon_{yy}$, respectively) within the 3 ML and 14 ML samples, respectively. (g)-(h) Values of $\varepsilon_{xx}$ and $\varepsilon_{yy}$, respectively, integrated across the region marked by the dashed black rectangle shown in (c). Along [111], the \LAO{} region in the 3 and 14 ML samples is subjected to expansion and contraction, respectively. However, along [1$\bar{1}$0], strain is unresolved.}
\label{FIG3}
\end{figure*}

To elucidate the enhancement of superconducting properties of the 3 ML (111) \STOLAOalumina{} compared with the bare 14 ML (111) \STOLAO{} samples, we mapped and analyzed the strain in the samples employing the geometric phase analysis (GPA) method to the HAADF STEM images. Summarized in Figure \ref{FIG3}, GPA results indicate that along the growth direction, [111], the \LAO{} region in the 3 ML sample expands with respect to the \STO{} substrate. Conversely, an opposite behavior is observed for the 14 ML sample. However, along the perpendicular direction, [1$\bar{1}$0], negligible strain values are calculated for both samples. Within the sensitivity of this measurement, we cannot resolve whether this strain is contraction or expansion. Furthermore, real-space measurements of interplanar distances corroborate these results, and show that the \LAO{} region in the 3 and 14 ML samples expand and contract along the [111] direction, respectively, by about 3 \%. In contrast, along the [1$\bar{1}$0] direction, both \STO{} and \LAO{} regions attain practically the same interplanar spacing. See Figures S4-8 and Table S1 for further details.

\section{Discussion}
In bulk samples of doped \STO{}, \Tc{} is increased by applying strain \cite{Ahadi2019, Herrera2019}. It has been shown that a thin layer of \LAO{}, below the critical thickness for conductivity, on [100] \STO{}, expands perpendicular to the interface. This expansion decreases when the \LAO{} layer thickness exceeds the critical thickness to allow charge transfer to the interface \cite{Cancellieri2011}.

The amorphous alumina, in that sense, allows conductivity with a very thin \LAO{} layer while the \LAO{} lattice is significantly expanded relative to the relaxed \LAO{} as we observe in our TEM GPA analysis for the (111) \STOLAOalumina{} interface. We conjecture that since the \LAO{} expands in a direction perpendicular to the interface, it exerts strain on the topmost \STO{} layer, where conductivity and superconductivity occur. This results in enhanced superconducting properties. Our description is consistent with the fact that the best superconducting properties are observed close to the critical thickness for conductivity both for the (111) and for the (100) \STOLAOalumina{} interfaces. Adding more \LAO{} layers, results in a lower \Tc{} as demonstrated in Table S1. Based on our analysis, we suggest that this decrease in T$_c$ is due to a smaller \LAO{} lattice expansion and consequently reduced strain.

Earlier structural analyses identified lattice expansion perpendicular to the interface in thin \LAO{} layers deposited on [100] \STO \cite{Cancellieri2011}. However, it is noteworthy that the maximum \Tc{} observed at the critical thickness in (100) interfaces is smaller than that in (111) interfaces (365mK and 420mK, respectively). 

What is the reason for the different \Tc? We utilize our findings for the (111) interface and reference the results from Ref.\cite{Cancellieri2011} for the (100) interface. We analyze the lattice expansion in the \LAO{} region in both cases. While a 3\% lattice expansion (relative to relaxed, bulk \LAO) is observed for the [111] direction, an order of magnitude smaller expansion of 0.3\% is found for the (100) sample. This observation corresponds with our findings of lower \Tc{} in (100) interfaces and further supports our claim that the enhanced superconductivity in (111) \STOLAOalumina{} originates from epitaxial strain.

Under our growth conditions, depositing alumina directly on \STO{} without \LAO{} is insufficient to generate neither conductivity nor superconductivity. Ryu et al. studied the uncapped (111) \STOLAO{} interface (without alumina) and compared a nonconducting, 3 ML \STOLAO{} sample with a conducting, 20 ML one \cite{Ryu2022}. They found electronic reconstruction without atomic reconstruction for both interfaces. In addition, the polar structure persists into the \LAO{} layer. An atomic reconstruction is avoided for both the conducting (20 ML in their case) and the nonconducting (3 ML) sample. We suggest that the absence of atomic reconstruction allowed by the very thin \LAO{} layer enables the appearance of conductivity in our 3 ML interface. Without the \LAO{} layer, the \STO{} surface reconstructs, giving a non-polar interface, which is probably the cause for the lack of conductivity.  

We also note that the data in Figure \ref{FIG2} suggest that while at 300K the Hall number increases with \LAO{} thickness, the number of mobile carriers (inferred from the low temperature Hall number) at the interface is, in fact, the highest at the critical thickness. But since the maximum point of the superconducting dome is shifted to higher \Tc{} values, it is not enough to consider the carrier density, which can be tuned for all samples by gate voltage.

\section{Summary and conclusion}
We show that deposition of amorphous alumina on \STOLAO{} interfaces reduces the critical thickness to be: two unit cells for the (100) interface and 3 ML for the (111) interface. The superconducting \Tc{} is enhanced for the (111) interface at the critical thickness of 3 ML. A detailed structural analysis by TEM leads us to believe that the \Tc{} enhancement is due to strain exerted by the expanded \LAO{} layer in the vicinity of the interface. Our results are important for improving superconducting devices from polar oxide interfaces. 

\section{Methods}

\subsection{Sample Preparation}
\textbf{Sample Deposition.} \LAO{} was deposited on atomically flat \STO{} substrates using the pulsed laser deposition (PLD) technique following Ref. \cite{Rout2017}. The deposition was monitored by reflection high energy electron diffraction (RHEED). A successive layer of 10nm of amorphous alumina was deposited at room temperature \textit{in situ} PLD using a sapphire single crystal target, at oxygen pressure of 1.5mTorr and laser fluence of 11.5mJ/mm$^2$, or via \textit{ex-situ} atomic layer deposition (ALD) \cite{Miron2019}. The results shown throughout this paper are independent of the alumina deposition method. 

\textbf{TEM sample preparation.} The 3 ML sample was prepared by a conventional cross section method, including mechanical polishing using a MultiPrep™ system (Allied High Tech, USA) followed by Ar ion milling using a PIPS II (Gatan, USA) apparatus at energies ranging from 4 down to 0.2 keV. To minimize damage induced by the ion milling, the sample was cooled to a temperature of 193 K. The 14 ML sample was prepared by a Ga$^+$ focused ion beam (FIB) Helios™ 5 UC DualBeam (Thermo Fisher Scientific, USA) device employing the lift-out technique, using energies from 30 down to 2 keV. To protect the samples from ion beam damage, amorphous carbon and tungsten were deposited before the milling process.

\subsection{TEM characterization}
Imaging and compositional analysis were carried out using an aberration corrected scanning transmission electron microscope (STEM) Titan Themis G2 60-300 (Thermo Fisher Scientific, USA) operated at 300 keV and equipped with a Dual-X detector (Bruker, USA) having a collection solid angle of 1.76 srad. High-angle annular dark-field (HAADF) Z-contrast imaging with drift corrected frame integration function was used to characterize the crystal structures employing an angular collection range of 60 to 200 mrad and a beam diameter below 1 Å. Energy-dispersive X-ray spectroscopy (EDS) was used to map and quantify elements based on the Sr L, Ti K, La L, and Al K characteristic energies; data analysis was carried out in Velox software (Thermo Fisher Scientific, USA). The elemental maps were pre-filtered using 3 px average, background corrected using a multi-polynomial fit (background order: 1 sloped), and post-filtered using Radial Wiener filter (highest frequency: 80, edge smoothing: 5). Concentration profiles were obtained by quantification without implementing absorption correction and post-filtering. The strains within the samples were extracted from HAADF STEM images by the geometric phase analysis (GPA) \cite{Hytch1998} method using Strain++ program \cite{Peters2024}, with reciprocal lattice vectors $g_{1\bar{1}0}$ and $g_{111}$ and a mask size of $g_{1\bar{1}0}$.

\subsection{Transport Measurements.} 
\textbf{Hall Measurements}. The Hall number was measured using a Keithley 6221 current source and Keithley 2182A nanovoltmeter using a Quantum Design PPMS. The Hall number is inferred from the linear Hall resistivity measured in a Van-der-Pauw configuration, after averaging two contact configurations and anti-symmetrizing with respect to the applied magnetic field. 

\textbf{Superconducting Properties}. The superconducting properties were measured using standard ac lock-in technique (Stanford SR830, frequencies in the range of 20 to 30Hz) inside an Oxford Instruments Triton 400 dilution refrigerator. The critical temperature is defined as the temperature in which the resistance is half of the normal state resistance at 0.7K.
\begin{acknowledgments}
The experimental work performed at Tel Aviv University was supported by the Pazy Research Foundation Grant No. 326-1/22, Israel Science Foundation (ISF) Grants No. 3079/20 and 476/22, the Tel Aviv University Quantum Research Center and the Oren Family Chair for Experimental Physics. A. K. acknowledges support by the ISF Grant No. 2973/21. Work at the Technion was funded by the ISF Grant No. 1351/21, with support from the Micro-
Nano Fabrication \& Printing unit (MNF\&PU).
\end{acknowledgments}

\bibliography{STOLAOAL2O3} 

\end{document}